\documentclass[a4paper]{article}
\usepackage{mymacros}
\usepackage{epsfig}
\title{Renormalization-Scheme-Independent Perturbation Theory
	by Resumming Logarithms}
\author{Chris Dams\footnote{C.Dams@science.ru.nl},
	Ronald Kleiss\footnote{R.Kleiss@science.ru.nl}\\
	\vbox{\hsize0.9\hsize\vskip 1ex
	\small Institute for Theoretical Physics\break
	Radboud University\break
	Toernooiveld 1\break
	6525 ED Nijmegen\break
	The Netherlands}}
\date{December 17, 2004}
\begin{document}

\maketitle

\begin{abstract}
Results of perturbation theory in quantum field theory generally depend
on the renormalization scheme that is in use. In particular, they depend
on the scale. We try to make perturbation theory scheme invariant by
re-expanding with respect to a scheme invariant quantity. Furthermore, we
investigate whether the potentially large logarithms in such an expansion
cause inaccuracy and how this can be improved.
\end{abstract}

\section{Introduction}
The occurrence of divergencies in perturbative quantum field theory has
made it clear that although measurable quantities should be finite, this need
not be the case for the parameters of the theory. To handle this,
divergencies are regularized. To obtain an order-by-order finite
perturbation expansion, it should be specified what to do with
divergencies as they occur at each loop level. This introduces an arbitrary
constant for each loop level. Although the full perturbation series should
not depend on these constants, the truncated one does. This is a problem
if one is trying to approximate a physical quantity that, by the definition
of ``physical quantity'', should not depend on these unphysical parameters.

It has been pointed out in~\cite{stevenson} that scheme invariant
quantities can be constructed out of the scheme-dependent ones. Therefore, the
natural thing to do would be to try to rewrite the perturbation series
into a series expansion with respect to the quantity~$X_1$, the
invariant that can be calculated at the one loop level.
Actually, that would be a series expansion
in~$1/X_1$, since $1/X_1$ is the small quantity if we are in the perturbative
regime.

An expansion in $1/X_1$ will, however, contain logarithms in the expansion
coefficients. At high energies, these logarithms will dominate all
other contributions to the expansion coefficients. We will resum these
logarithms and investigate whether this gives more reliable results.

Several attempts to remedy this problem of renormalization scheme 
dependence have been proposed. They all suffer from a great deal of
arbitrariness and/or the mathematics involved is more complicated
than that of simply manipulating power series, as we will be doing.

To obtain our results, we used the simple case of a one-parameter
theory. We did not yet consider more complicated cases, but will do
this later. When doing explicit calculations, we use QCD with five
massless quarks, as a concrete example of a one-parameter theory.

\section{Renormalization Scheme Invariants}
\label{sec:invars}
In this section we outline how renormalization scheme-independent quantities
can be combined into scheme invariant ones. This is also explained
in \cite{stevenson}~and \cite{maxwell}. For self-containedness of this
paper and also to make clear what our conventions are, we will repeat
this in this section. The coupling constant will be denoted
by~$a$. The key idea is that the consistency of perturbation theory
requires that a result up to terms of order~$a^n$ should only differ from
the exact answer up to terms of order~$a^{n+1}$. If we consider the
case $n=1$, we see that the running of the
coupling makes the use of $a$ as an expansion parameter scheme
dependent. This indeed is an effect that starts at order~$a^2$.
I.e., the lowest
order term in the beta function is of order~$a^2$. 

The renormalization scheme can be specified by giving the renormalization
scale~$s$ and the scheme-dependent beta function
coefficients~$\beta_2,\beta_3,\beta_4,\ldots$~. The coupling constant
depends on the scale via
\beq
\frac{\partial a}{\partial s}=\beta(a)=\beta_0a^2+\beta_1a^3+\beta_2a^4+\cdots,
\eeq
where $ds=d\mu/\mu$ with $\mu$ the mass scale used in dimensional
regularization. This differential equation can be integrated to give
\beq\label{eq:running}
s=-\frac1{\beta_0a}
   +\frac{\beta_1}{\beta_0^2}\log\frac{\beta_0+\beta_1a}{\beta_1a}
   +\int_0^ada'\,
		\left(\frac1{\beta(a')}-\frac1{\beta_0(a')^2+\beta_1(a')^3}\right).
\eeq
This solution implies the choice of a boundary value. Our choice is
similar to the one in~\cite{stevenson}. Taking the set
of variables that consists of~$s$ and the scheme-dependent beta function
coefficients as independent, we can derive that
\beq
\frac{\partial a}{\partial\beta_i}
	=\beta(a)\int_0^ada'\frac{(a')^{i+2}}{\beta(a')^2}.
\eeq

As an example, we imagine that we have calculated a physical quantity~$R$
up to fourth order. From this example, we hope, it will be clear how
this can be generalized to arbitrary order. Having calculated $R$ up to 
fourth order means that we have
\beq\label{eq:Rina}
R\sim r_0a+r_1a^2+r_2a^3+r_3a^4.
\eeq
Consistency of perturbation theory requires that
\beq
\frac{dR}{d\,\textrm{scheme}}\sim a^5.
\eeq
To be more concrete, independence of~$s$ implies that
\beq
\left(r_0+2r_1a+3r_2a^2+4r_3a^3\right)\frac{\partial a}{\partial s}+
	\frac{\partial r_1}{\partial s}a^2+\frac{\partial r_2}{\partial s}a^3+\frac{\partial r_3}{\partial s}a^4\sim a^5.
\eeq
From this, equations for second, third and fourth order can be extracted.
Substituting the beta function for $\partial a/\partial s$, we find
\beq
\eqalign{
r_0\beta_0+\frac{\partial r_1}{\partial s}&=0;\cr
2\beta_0r_1+r_0\beta_1+\frac{\partial r_2}{\partial s}&=0;\cr
3\beta_0r_2+2\beta_1r_1+r_0\beta_2
	+\frac{\partial r_3}{\partial s}&=0.\cr
}
\eeq
To obtain the equations that follow from independence of $\beta_2$, we
expand $\partial a/\partial \beta_2$ up to fourth order. We have
\beq
\frac{\partial a}{\partial\beta_2}=\frac1{\beta_0}a^3+O(a^5).
\eeq
Using this and then demanding that the third and fourth order of
$\partial R/\partial\beta_2$ are zero, we get the equations
\beq
\eqalign{
\frac{r_0}{\beta_0}+\frac{\partial r_2}{\partial\beta_2}&=0;\cr
2\frac{r_1}{\beta_0}+\frac{\partial r_3}{\partial\beta_2}&=0.\cr
}
\eeq
Finally, independence of $\beta_3$ gives the equation
\beq
\frac{r_0}{2\beta_0}+\frac{\partial r_3}{\partial\beta_3}=0.
\eeq
Integrating the equations for $r_1$,~$r_2$, and~$r_3$, we find
\beq\label{eq:rsol}
\eqalign{
r_1&=X_1-\beta_0r_0s;\cr
r_2&=X_2+\frac{r_1^2}{r_0}-\frac{r_0\beta_2}{\beta_0}
	+\frac{\beta_1r_1}{\beta_0};\cr
r_3&=X_3+\frac{3r_1X_2}{r_0}+\frac52\frac{\beta_1r_1^2}{\beta_0r_0}
	-\frac{r_0\beta_3}{2\beta_0}-\frac{2r_1\beta_2}{\beta_0}
	+\frac{r_1^3}{r_0^2},\cr
}
\eeq
where the quantities~$X_i$ are the renormalization scheme invariants.
They arise because the values of the $r_i$ are obtained from differential
equations and hence need constants of integration. The values of the
$X_i$ can be calculated from the values of $r_i$ and $\beta_i$ once
these have been calculated. It should be made sure that the $r_i$ and
$\beta_i$ have been calculated in the same renormalization scheme, of
course, and then it will turn out that the values of $X_i$ no longer
depend on the scheme that was used to obtain $r_i$ and $\beta_i$.

The acute reader will have noticed that
we actually expressed $r_2$ and $r_3$ into $r_1$ instead of in
$s$. Hence, we actually label our renormalization scheme by
$r_1,\beta_2,\beta_3,\ldots$ rather than by the set of parameters that we
mentioned earlier.
The reason that we do this is that if we are going to re-expand
into~$1/X_1$ we would rather have $r_1$ than~$s$ in the expansion coefficients 
because in a suitable renormalization scheme $r_1$ will be a number
of order unity while $s$ will be of order $1/a$, which is large.
In the end it will
hopefully turn out that the result will not depend on what kind of ``suitable
renormalization scheme'' we used, however, we are not yet at this point so
we should still make sure that we do not have large expansion coefficients.

\section{Expansion in $1/X_1$}
For an expansion to work, we need to know that we are expanding with
respect to a small parameter. The idea of perturbation theory is that
the coupling constant, $a$, is a small quantity. Equation~(\ref{eq:running})
can be expanded in powers of~$a$. We have
\beq\label{eq:sina}
\eqalign{
s&\sim
-\frac1{\beta_0a}
+\frac{\beta_1}{\beta_0^2}\log\left(\frac{\beta_0}{\beta_1a}\right)
+\frac{\beta_1^2a}{\beta_0^3}
-\frac{\beta_2a}{\beta_0^2}
-\frac{\beta_1^3a^2}{2\beta_0^4}
-\frac{\beta_3a^2}{2\beta_0^2}
+\frac{\beta_1\beta_2a^2}{\beta_0^3}\cr
&\qquad-\frac{\beta_4a^3}{3\beta_0^2}
+\frac{\beta_2^2a^3}{3\beta_0^3}
-\frac{\beta_2\beta_1^2a^3}{\beta_0^4}
+\frac{2\beta_3\beta_1a^3}{3\beta_0^3}
+\frac{\beta_1^4a^3}{3\beta_0^5}.\cr
}
\eeq
From this series we see that if $a$ is small, $s$ must be large. Indeed,
in the case of QCD, $s$ is given by
\beq
s=\log\left(\frac{\mu^2}{\Lambda_{\textrm{\scriptsize QCD}}^2}\right),
\eeq
where $\mu$ is the renormalization scale as it occurs in dimensional
regularization and $\Lambda_{\textrm{\scriptsize QCD}}$ is of the order
of the
energy range where the coupling is large and perturbation theory is
not accurate. Therefore, the perturbation series can also be written
as an expansion in $1/s$. For this we use the inverse of the
expansion~\ref{eq:sina}. We have
\beq\label{eq:ains}
\eqalign{
a&\sim
-\frac1{\beta_0s}
+\frac{\beta_1}{\beta_0^3s^2}L_s
+\frac{\beta_1^2}{\beta_0^5s^3}
-\frac{\beta_2}{\beta_0^4s^3}
-\frac{\beta_1^2}{\beta_0^5s^3}L_s
-\frac{\beta_1^2}{\beta_0^5s^3}L_s^2\cr
&\qquad-\frac{\beta_1^3}{2\beta_0^7s^4}
+\frac{\beta_3}{2\beta_0^5s^4}
-\frac{2\beta_1^3}{\beta_0^7s^4}L_s
+\frac{3\beta_2\beta_1}{\beta_0^6s^4}L_s
+\frac{\beta_1^3}{\beta_0^7s^4}L_s^3
+\frac{5\beta_1^3}{2\beta_0^7s^4}L_s^2,\cr
}
\eeq
where $L_s=\log(-\beta_1/(\beta_0^2s))$.

Now we assume that a suitable
renormalization scheme has been chosen, and hope to obtain results that
turn out not to depend on our ``suitable renormalization scheme'' and hence
might also have been calculated if we had started out with an unsuitable
renormalization scheme. In a suitable renormalization scheme, we expect
that expansion coefficients are not large. In particular $r_1$ is
expected to be of order unity. From equation~(\ref{eq:rsol}) it follows
that $X_1=r_1+\beta_0r_0s$. We conclude that $X_1$ must
be a large quantity, because $s$ is. Hence, the expansion
\beq\label{eq:sinX}
\frac1s=\frac{\beta_0r_0}{X_1-r_1}
	\sim
	\frac{\beta_0r_0}{X_1}
	+\frac{\beta_0r_0r_1}{X_1^2}
	+\frac{\beta_0r_0r_1^2}{X_1^3}
	+\frac{\beta_0r_0r_1^3}{X_1^4} 
\eeq
is a good expansion. We substitute this equation into
equation~(\ref{eq:ains}) obtaining an expansion of $a$ in $1/X_1$.
This expansion is substituted into equation~(\ref{eq:Rina}). We then
obtain an expansion of the physical quantity~$R$ into $1/X_1$.
Furthermore, in this expansion, we substitute for $r_2,r_3,r_4,\ldots$
the values as given by equation~(\ref{eq:rsol}).
It should be noted that although $X_1$ does not depend on the
renormalization scheme, it is dependent upon the physical quantity
under consideration. There is, however, nothing wrong with using
a different expansion parameter for every different physical quantity.
The expansion that is obtained by making all these substitutions is
\beq\label{eq:RinX}
\eqalign{
R&\sim
-\frac{r_0^2}{X_1}
+\frac{r_0^3\beta_1}{\beta_0X_1^2}L_X
-\frac{r_0^3X_2}{X_1^3}
+\frac{r_0^4\beta_1^2}{\beta_0^2X_1^3}(1-L_X-L_X^2)\cr
&\qquad+\frac{r_0^4X_3}{X_1^4}
+\frac{3r_0^4\beta_1X_2}{\beta_0X_1^4}L_X
+\frac{r_0^5\beta_1^3}{\beta_0^3X_1^4}
	\left(\textstyle -\frac12-2L_X+\tfrac52L_X^2+L_X^3\right),\cr
}
\eeq
where $L_X$ is given by $L_X=\log(-r_0\beta_1/(\beta_0X_1))$.
We see that in this expansion all scheme-dependent terms have
canceled. These scheme-dependent terms are the ones involving~$r_1$ and
$\beta_i$ with $i\geq2$.
Because of the cancellation of these terms, we have
obtained a scheme-independent perturbation series.

For this method to work for any order in perturbation theory, we
must prove that the cancellation of scheme-dependent terms happens at
every order and not just up to fourth order. For this purpose, note
that our expansion is an expansion with respect
to the variables
$1/X_1,\allowbreak1/\beta_0,\allowbreak\beta_1,\allowbreak
\beta_2,\allowbreak\beta_3,\allowbreak\ldots,\allowbreak r_0,\allowbreak
r_1,\allowbreak X_2,\allowbreak X_3,\allowbreak X_4,\allowbreak\ldots,L_X$.
If we refer to the order of a term in the series, we mean the order in
$1/X_1$. We should prove that actually the variables $r_1,\allowbreak
\beta_2,\allowbreak\beta_3,\allowbreak\ldots$ do not occur in this
series expansion. Let us assume that such a variable actually does
occur at some order $n$ in the expansion.

We introduce~$v$ as an alias for one of the offending variables
(there might be several offending variables) that occurs at order $n$.
This means that $\partial R/\partial v$ is of order~$n$. We consider
what happens if we re-expand the expression for $R$ in $a$ again,
and then differentiate with respect to~$v$. To do this we first have
to expand $1/X_1=1/(\beta_0r_0s+r_1)$ in $1/s$ and then use
equation~(\ref{eq:sina}) to expand this in $a$ again. $X_1$ was
defined in such a way that it actually does not depend on scheme-dependent
quantities
such as~$v$. Therefore, we know that the entire series of
$1/X_1$ in $a$ does not depend on $v$. Furthermore, we note that
during re-expanding and differentiating the order of a term in $1/X_1$,
$1/s$ or $a$, whichever applies, never decreases. Therefore, all the
invariant terms up to
order $n$ after differentiation cause terms that are at least of order
$n+1$.  The only terms that can give a contribution of order $n$ are the
scheme-dependent terms. However, the terms obtained by differentiating $a$,
using the chain rule, are of higher order, so this does not contribute.
The conclusion is that the derivative with respect to~$v$ up to order~$n$
is the same before re-expanding as after re-expanding, except that
$1/X_1$ is to be replaced by $-a/r_0$ and $L_X$ is to be replaced by
$\log(\beta_1a/\beta_0)$. Hence, the physical quantity~$R$ up to
order~$n$ depends on the scheme at order~$n$. This is a contradiction
with the starting point of section~\ref{sec:invars}. Therefore the
expansion in~$1/X_1$ must have renormalization scheme independent
coefficients.

\section{Resumming the Logarithms}
In this section, we will resum all logarithms~$L_s$ that occur in
equation~(\ref{eq:ains}). We start out by observations that have
been made from computer algebra experimentation,
but in the end we will prove our
results to be correct to all orders. By looking at the
expansion~(\ref{eq:ains}), we observe that the leading logarithms, i.e.,
terms of the order $L_s^{n-1}/s^n$ can be summed into the quantity~$1/s'$
defined by
\beq
\frac1{s'}=\frac1s\frac1{1+\beta_1L/(\beta_0^2s)}.
\eeq
After this resummation we have the expansion
\beq
\eqalign{
a(s',L)&\sim
-\frac{1}{\beta_0s'}
-\frac{\beta_2}{\beta_0^4(s')^3}
+\frac{\beta_1^2}{\beta_0^5(s')^3}(1-L)
+\frac{\beta_1^3}{\beta_0^7(s')^4}
	\left(\textstyle-\frac12+L-\frac{L^2}2\right)\cr
&\qquad+\frac{\beta_3}{2\beta_0^5(s')^4}
+\frac{\beta_1^4}{\beta_0^9(s')^5}
	\left(\textstyle-\frac76+2L-\frac12L^2-\frac13L^3\right)\cr
&\qquad+\frac{3\beta_1^2\beta_2}{\beta_0^8(s')^5}(1-L)
-\frac{5\beta_2^2}{3\beta_0^7(s')^5}
+\frac{\beta_1\beta_3}{6\beta_0^7(s')^5}
-\frac{\beta_4}{3\beta_0^6(s')^5}\cr
}
\eeq
Looking at the highest order logarithms in this expansion, we recognize the
expansion of $\beta_1/(\beta_0^3s^2)\log(1-\beta_1L/(\beta_0^2s'))$.
We therefore define a quantity $L'$ by
\beq
L'=\log\left(1-\frac{\beta_1}{\beta_0^2s'}L\right).
\eeq
It now turns out that after rewriting the expansion for~$a$ with respect
to $1/s'$ and $L'$, we recover the original expansion where $1/s$ has
been replaced by~$1/s'$ and $L$ by~$L'$. Because of this we can
iterate this procedure and obtain a sequence of values $1/s_n$ and
$L_n$ from the iteration
\beq\label{eq:iter}
\eqalign{
\frac1{s_{n+1}}&=\frac1{s_n}\frac1{1+\frac{\beta_1}{\beta_0^2s_n}L_n};\cr
L_{n+1}&=\log\frac1{1+\frac{\beta_1}{\beta_0^2s_n}L_n},\cr
}
\eeq
where we have rewritten $L_{n+1}$ in terms of $s_n$~and $L_n$ instead
of in terms of $s_{n+1}$~and~$L_n$. We note that this iteration increases
the order of $L$ with respect to $1/s$. Therefore, in the perturbative
regime, the iteration should make $L$ converge to zero. This resums all
logarithms into a new value for $1/s$.

A curve can be drawn through the sequence of points~$(1/s_n,L_n)$.
This curve is given by
\beq
\eqalign{
\frac1{s(x)}&=\frac1{s_\infty}\frac1{1+\frac{\beta_1}{\beta_0^2s_\infty}x};\cr
L(x)&=\log\frac1{1+\frac{\beta_1}{\beta_0^2s_\infty}x}-x,\cr
}
\eeq
where $x$ parameterizes the curve and different curves (for different
initial values of $1/s$~and~$L_s$) are labeled by~$1/s_\infty$. That this is
correct, can be seen by checking that the iteration for the pair~$(1/s,L)$
is recovered if $x$ is iterated using the prescription
\beq
x_{n+1}=\log\frac1{1+\frac{\beta_1}{\beta_0^2s_\infty}x_n}.
\eeq
We see that also for the $x_n$ the property holds that $x_{n+1}$ is
of higher order than $x_n$, hence in the perturbative regime it
should converge to zero. In this case we note that
$1/s(x)\to1/s_\infty$, which explains our notation~``$1/s_\infty$'' to label
the different curves.

We must still prove our assertion that the iteration preserves the expansion
to all orders in~$1/s$.  It is sufficient to show that the
value of~$a$ is constant along the curves introduced above.
The infinitesimal form of the curves is
\beq\label{eq:symm}
\eqalign{
\delta s&=\frac{\beta_1}{\beta_0^2}\delta x;\cr
\delta L_s&=\left(-1-\frac{\beta_1}{\beta_0^2 s}\right)\delta x\cr
}
\eeq
Proving that this is a symmetry of the expansion~(\ref{eq:ains}) is
also sufficient to see that the iteration works. This is what we will do
in the next section.

\section{Proof of the Iteration}
Here we will prove that the transformation~(\ref{eq:symm}) is a
symmetry of the expansion equation~(\ref{eq:ains}). This, at the
same time, shows that the iteration of equation~(\ref{eq:iter}) works
to all orders and that the logarithms in the expansion can be made
zero. We introduce a quantity~$\bar a$ that is
a power series in $1/s$ and $L_s$. This quantity has the definition
\beq\label{eq:adiffeq}
\eqalign{
\frac{\partial\bar a}{\partial(1/s)}
	&=-\left(s^2+\frac{\beta_1s}{\beta_0^2}\right)\beta(\bar a);\cr
\frac{\partial\bar a}{\partial L_s}&=\frac{\beta_1}{\beta_0^2}\beta(\bar a);\cr
	\left[s\bar a(1/s,L_s=0)\right]_{s\to\infty}&=-1/\beta_0.\cr
}
\eeq
We will show that this quantity~$\bar a$ is actually the same as~$a$.
First note that these differential equations are consistent because
$\partial/(\partial(1/s))$ and $\partial/\partial L_s$ commute. Secondly,
if we confine ourselves to the surface $L_s=\log(-\beta_1/(\beta_0^2s))$
we have, as is verifiable by using the chain rule for differentiation,
$d\bar a/ds=\beta(\bar a)$, so, also using the boundary condition, we
see that on this surface $a$ and $\bar a$ are the same.
We now show that for every order in $1/s$,
there is a finite number of terms. This ensures that also away from the
surface~$L_s=\log(-\beta_1/(\beta_0^2s))$ these functions are the same,
because it is impossible to express
$s$ and $L_s$ into each others in a finite number of terms. The pre-factors
of the expansion in $1/s$~and $L_s$ of $\bar a$ can be obtained by setting
$\bar as=-1/\beta_0$ in the derivatives
$\partial^{n+m}a/(\partial L_s^m\partial (1/s)^n)$. At first sight, it
may seem that problems could be caused by terms of the form $\bar a^ms^n$
where $n>m$. However, the fact that the differential equations have
a solution that is an expansion in $L_s^m/s^n$ with $n>0$~and~$m\geq0$,
ensures that these problematic contributions will cancel if we
substitute $\bar as\to-1/\beta_0$. Still, before this substitution
is made, there will be terms of the form $\bar a^p(1/s)^{p+q}$, with
$p>0$~and~$q\geq 0$. If we
consider the quantity $d^n\bar a/d (1/s)^n$, the maximum
value of $q$ for which this type of monomial will occur is equal to
$n-1$. Differentiations with respect to $L_s$ increase the order in~$a$,
hence, in $\partial^{n+m}a/(\partial L_s^m\partial (1/s)^n)$, the maximum
value for $q$ will be $n-1-m$. Hence, if $m\geq n$ we will only have
terms containing $\bar a^p(1/s)^q$ with $p>q$. If we substitute
$\bar as=-1/\beta_0$, taking $s\to\infty$, these terms will become zero
and thus do not 
contribute. From this we see that the maximum order in~$L_s$ that
occurs in the coefficient of $1/s^p$ in the expansion in $s$ is~$p$.
Hence, this coefficient of $1/s^p$ contains a finite number of terms,
as we set out to show. We conclude that the quantity $\bar a$
indeed has the same expansion with respect to $1/s$ and $L_s$ as $a$.

Furthermore, from equations (\ref{eq:symm})~and
(\ref{eq:adiffeq}), it follows that $da/dx=0$. Therefore the
symmetry~(\ref{eq:symm}) is indeed a symmetry of the expansion of $a$
in $1/s$~and~$L_s$, and the iteration in equation~(\ref{eq:iter}) keeps
the value of~$a$ constant to all orders in~$1/s$.

\section{Resumming $L_X$}
The symmetry of equations~(\ref{eq:symm}) can be turned into a symmetry
of the expansion of a physical quantity in  $1/X_1$ and $L_X$. This
will enable us to perform resummation of logarithms in such an expansion.
From $X_1=r_0\beta_0s+r_1$, it follows that
\beq\label{eq:Xsymm}
\delta X_1=r_0\beta_0(\delta s)=\frac{r_0\beta_1}{\beta_0}\delta x.
\eeq
Turning $L_s$ into $L_X$ is done via
\beq
L_s=L_X+\log\frac{X_1}{X_1-r_1}.
\eeq
Applying $\delta$ on both sides gives
\beq
\left(-1-\frac{\beta_1}{\beta_0s}\right)\delta x
	=\delta L_X-\frac{r_1}{X_1}\delta X_1.
\eeq
Equation~\ref{eq:Xsymm} then gives
\beq\label{eq:LXsymm}
\delta L_X=-\left(1+\frac{r_0\beta_1}{\beta_0X_1}\right)\delta x.
\eeq
Using this symmetry it is possible to turn $L_X$ into zero, thereby
ridding ourselves of logarithms. The value of $X_1$ that is obtained while
turning $L_X$ to zero will be called $\tilde X_1$. Integrating equations
(\ref{eq:Xsymm})~and~(\ref{eq:LXsymm}), we obtain the equation
\beq\label{eq:Xeq}
X_1=\tilde X_1
	-\frac{r_0\beta_1}{\beta_0}\log\frac{-r_0\beta_1}{\beta_0\tilde X_1}.
\eeq
This can be expressed in the Lambert $W$-function. This function is
by definition the solution to $W(z)e^{W(z)}=z$. We have
\beq
\tilde X_1=\frac{r_0\beta_1}{\beta_0}
	W\left(-e^{\frac{\beta_0X_1}{r_0\beta_1}}\right).
\eeq
Hence, the conclusion is that we turned the standard perturbation theory
into an expansion in the quantity~$1/\tilde X_1$. The expansion looks
as displayed in equation~(\ref{eq:RinX}) with $X_1$ replaced
by~$\tilde X_1$ and all terms that have an $L_X$ removed. Since this
reduces the number of terms considerably, let us display a few more.
We have
\beq
\eqalign{
R&\sim
-\frac{r_0^2}{\tilde X_1}
+\frac{r_0^4\beta_1^2}{\beta_0^2\tilde X_1^3}
-\frac{r_0^3X_2}{\tilde X_1^3}
-\frac{r_0^5\beta_1^3}{2\,\beta_0^3\tilde X_1^4}
+\frac{r_0^4X_3}{\tilde X_1^4}
-\frac{7\,r_0^6\beta_1^4}{6\,\beta_0^4\tilde X_1^5}
+\frac{3\,r_0^5\beta_1^2X_2}{\beta_0^2\tilde X_1^5}
-\frac{r_0^5X_4}{\tilde X_1^5}\cr
&\qquad+\frac{17\,r_0^7\beta_1^5}{12\,\beta_0^5\tilde X_1^6}
-\frac{3\,r_0^6\beta_1^3X_2}{2\,\beta_0^3\tilde X_1^6}
-\frac{4\,r_0^6\beta_1^2X_3}{\beta_0^2\tilde X_1^6}
+\frac{r_0^6X_5}{\tilde X_1^6}.\cr
}
\eeq

\section{Determining $\Lambda_{\textrm{\scriptsize QCD}}$}

\vskip-1ex
In their \textsl{Review of Particle Physics}~\cite{review}
the Particle Data Group
suggests using equation~(\ref{eq:ains}) to define
$\Lambda_{\textrm{\scriptsize QCD}}$. To be fully accurate, their
definition is not completely the same. Instead of our
$L_s=\log(-\beta_1/(\beta_0^2s))$ the PDG uses $L_s=-\log s$. This
amounts to a shift in the parameter~$s$. A way to see this is from
equation~(\ref{eq:running}). Adding a factor $-\beta_1/\beta_0^2$ inside
the logarithm in this equation, turns our expansion into the one of the
PDG\@. This is equivalent to adding a constant to the right-hand side
of this equation. This constant can then be moved to the left-hand side,
so we see that $s$ is indeed shifted.
The consequence is that
the $\Lambda_{\textrm{\scriptsize QCD}}$ that we use differs by
a multiplicative constant from the one of the PDG\@. We have
\beq\label{eq:lambda}
\Lambda_{\textrm{\scriptsize QCD}}^{\textrm{\scriptsize PDG}}
	=\Lambda_{\textrm{\scriptsize QCD}}^{\textrm{\scriptsize OURS}}
		\left(-\frac{\beta_0^2}{\beta_1}\right)^{\beta_1/(2\beta_0^2)},
\eeq
where the beta function coefficients are given in our conventions.
In the rest of the paper we use our conventions, hence we will
be writing $\Lambda_{\textrm{\scriptsize QCD}}$ for
$\Lambda_{\textrm{\scriptsize QCD}}^{\textrm{\scriptsize OURS}}$
and never mention
$\Lambda_{\textrm{\scriptsize QCD}}^{\textrm{\scriptsize PDG}}$ again.
Note however, that the expansion with respect to $1/X_1$~and~$L_X$
becomes different if we choose a different prefactor inside the
logarithm. We will see in the next section that using the symmetry
of equations (\ref{eq:Xsymm})~and~(\ref{eq:LXsymm}) resolves this
ambiguity.

\begin{table}
\centerline{
\begin{tabular}{|c|c|c|}
\hline
highest order & not using symmetry & using symmetry \cr
\hline
$1/s$ & \phantom091.5 & \phantom091.5\cr
$1/s^2$ & 282.2 & 249.0 \cr
$1/s^3$ & 247.5 & 249.3 \cr
$1/s^4$ & 248.6 & 249.0 \cr
\hline
\end{tabular}}
\caption{Values for $\Lambda_{\textrm{\scriptsize QCD}}$ in
MeV obtained by solving equation~(\ref{eq:sina}) to various orders
numerically, while using or not
using the symmetry in equation~(\ref{eq:symm}).
We used a number of quarks equal to five to obtain this result.}
\label{tab:Lambda}
\end{table}

The suggestion of the PDG to use equation~(\ref{eq:ains}) to define
$\Lambda_{\textrm{\scriptsize QCD}}$ is not very practical. In the
first place it would seem to be easier to use
equation~(\ref{eq:running}). Secondly, if one is going to use
equation~(\ref{eq:ains}), the symmetry from equation~(\ref{eq:symm}) is
useful to obtain a series that converges faster by transforming
$L_s$ to zero. The PDG gives $\alpha_s(M_Z)\approx0.1187$. In
table~\ref{tab:Lambda} we compare the value obtained for
$\Lambda_{\textrm{\scriptsize QCD}}$ using the symmetry and not using
it. The used beta function coefficients were calculated in~\cite{bfun}
and recently confirmed in~\cite{bfun2}.
The number of flavours was set to five. We indeed see faster convergence.

Below we will be using
$\Lambda_{\textrm{\scriptsize QCD}}$. The value that we use comes
from equation~(\ref{eq:ains}) to fourth order where we use our
symmetry to get rid of the $L_s$'s. If one uses standard
perturbation theory, one has to pick a suitable value for $\mu$ and
specify the renormalization scheme. In that case, the consistent way
to proceed is to
only use the beta function up to the loop level to which the rest of the
calculation is done.
The difference with our case is that while in the standard approach
$\alpha_s(M_Z)\approx0.1187$ appears as a fundamental constant,
in our approach $\Lambda_{\textrm{\scriptsize QCD}}$ would be the
fundamental quantity.
Determining the value of a fundamental constant from another one
is better done with as much accuracy as possible. This is the reason
that we use the beta function up to four loop level.

\section{How Invariant is Invariant, Really?}
Our method attempts to remedy the arbitrariness in choosing
the renormalization prescription, so we should now ask the question
to what extend this method itself is arbitrary. A
first possible source of arbitrariness is the choice of variables
to parameterize the renormalization prescription. We choose
the set of parameters $r_1,\beta_2,\beta_3,\beta_4,\ldots$ to
parameterize the renormalization prescription. If one chooses
a different set of variables, $\bar r_1=r_1+\Delta_1,
\bar\beta_2=\beta_2+\Delta_2, \bar\beta_3=\beta_3+\Delta_3$,
where the $\Delta$'s are constants,
the integration constants~$X_i$ from section~\ref{sec:invars}
also become different. However, if we have $\Delta_1=0$, it turns
out that in the end the final expansion coefficients still have the
same value, so
this is not an arbitrariness of our method.

This shows that the small invariant quantity that we decide to use for
expansion is more
important than the precise definition of the other invariants.
The question that might arise is what would happen
if we would expand with respect to some arbitrary function
of~$X_1$ instead of with respect to~$X_1$.
We could, for instance, expand with respect to the sine of~$X_1$.
The possibility of expressing
the coupling constant in one scheme as a power series in the coupling
constant in another scheme is a possibility that has been mentioned in
literature, for instance in~\cite{stevenson}. Our point of view is that the
possibility to use an arbitrary power series is not a fundamental
arbitrariness of perturbation theory. Note that if it were, it would
apply to any perturbative method in any branch of physics. Small
quantities that are used for expansion should be the ones that
come naturally with the problem under consideration, not the ones that
can be used to show that any approximation method
can be made to give wrong answers. In field theory, the situation
is that no renormalization scheme is a priori better than
any other, and this ambiguity can be parameterized by using an arbitrary
power series in the coupling constant. This is the reason
arbitrary power series of the coupling
constant can be useful to consider. Considering the fact that the
first scheme invariant arises naturally from the demand that a physical
quantity should not depend on the scale, it does not make sense to 
consider arbitrary functions of this, perhaps with the exception of
a translation in de definition of~$X_1$. I.e., a non-zero value
of~$\Delta_1$.

If $\Delta_1\neq0$, we obtain a different invariant variable
on the one loop level, namely $ X'_1=X_1+\Delta_1$. If we were
to expand in $1/X'_1$, we would indeed obtain a different
expansion. Here, it appears, we have finally found arbitrariness.
However, we actually already mentioned this. An example of this
arbitrariness is given in the paragraph that contains
equation~(\ref{eq:lambda}). The prefactor
that we choose inside the logarithm~$L_s$ is
equivalent to choosing a value for~$\Delta_1$. Resumming
the logarithms yields the same result for $\tilde X_1$ and, hence, also
the same result for the physical quantity~$R$.

\begin{figure}
\centerline{\epsfig{file=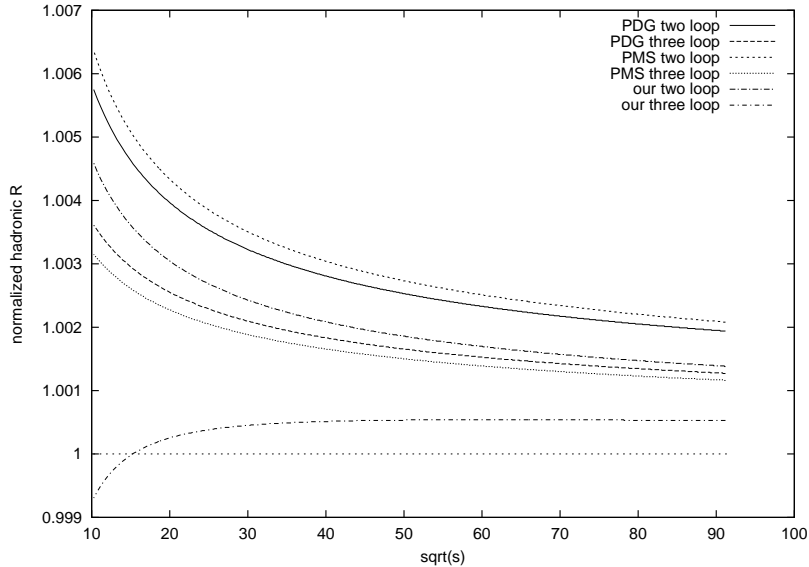,height=0.9\textwidth,angle=270}}
\caption{hadronic~$R$ as a function of the center of mass energy for
$N_f=5$, as found by various methods, normalized by dividing by the
PDG-one-loop result. Note that our method gives rather
different results from the PDG-method and the PMS-method.
From top to bottom we see graphs for PMS-two-loop, PDG-two-loop, [gap],
our-two-loop, PDG-three-loop, PMS-three-loop, [gap], and our-three-loop.}
\label{fig:hadrR}
\end{figure}

\begin{figure}
\centerline{\epsfig{file=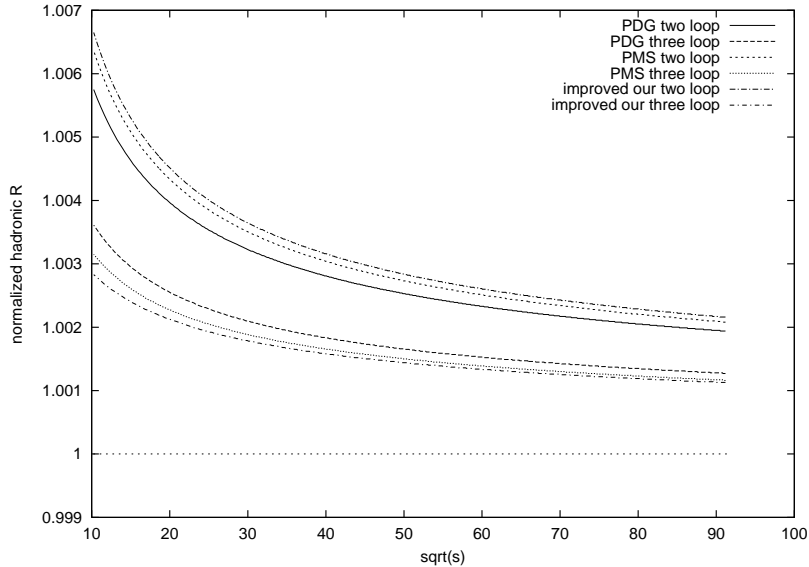,height=0.9\textwidth,angle=270}}
\caption{hadronic~$R$ as a function of the center of mass energy for
$N_f=5$, as found by various methods, normalized by dividing by the
PDG-one-loop result. As opposed to the previous graph,
our method now gives very similar results to the PDG-method and the
PMS-method.
From top to bottom we see graphs for
improved-our-two-loop, PMS-two-loop, PDG-two-loop, [gap], PDG-three-loop,
PMS-three-loop, improved-our-three-loop.}
\label{fig:imphadrR}
\end{figure}

\section{Hadronic-$R$}
\label{sec:hadrR}
In this section we consider massless QCD-corrections to
hadronic~$R_{\textscr{had}}$.
This quantity is by definition given by
\beq
R_{\textscr{had}}=\frac{\sigma(e^++e^-\to\textrm{hadrons})}{\sigma(e^++e^-\to\mu^++\mu^-)},
\eeq
where electroweak corrections are neglected.
The value of this quantity can be obtained up to third order from the review
paper~\cite{corr}. The beta function coefficients can be obtained from the
same paper or from~\cite{bfun}. $a$ is taken to be $\alpha_s/\pi$. For
the first renormalization scheme invariant~$X_1$, we have
\beq
\eqalign{
X_1&=r_1+\beta_0r_0s\cr
	&=\frac{365}{24}
   -11\zeta(3)
   -N_f\left(\frac{11}{12}-\frac{2}{3}\zeta(3)\right)
   +\left(-\frac{11}4+\frac16N_f\right)
		\log\left(\frac {s_{\textrm{\scriptsize CM}}}{
			\Lambda_{\textrm{\scriptsize QCD}}^2}\right),\cr
}
\eeq
where we have written $s_{\textrm{\scriptsize CM}}$ for the squared center of
mass energy, to avoid confusion with the quantity
$\log(\mu^2/\Lambda_{\textrm{\scriptsize QCD}}^2)$, that we also call~$s$.
The addition of $\beta_0r_0s$ has resulted in the replacement
$\mu\to\Lambda_{\textrm{\scriptsize QCD}}$, in $r_1$. Note that the 
value of $\Lambda_{\textrm{\scriptsize QCD}}$ is scheme dependent. For
instance, if we would use the MS-scheme (we are actually using
$\overline{\textrm{MS}}$), the two values of
$\Lambda_{\textrm{\scriptsize QCD}}$ would differ by a multiplicative constant.
The scheme
invariant is independent of this choice. A practical way to
proceed is to start from the expansion of $R_{\textscr{had}}$
up to and including terms of order $a^3$, substitute, equation~(\ref{eq:ains})
into it and expand up to terms of order $1/s^3$. In this we substitute
equation~(\ref{eq:sinX}) and expand again with respect to~$1/X_1$ up to
terms of order~$1/X_1^3$. Using computer
algebra this is a more or less trivial thing to do.
This cancels all dependence on the scheme, as we have argued. In particular,
we observe that results no longer depend on the renormalization
mass~$\mu$.

We plot a normalized variant of hadronic~$R$ as calculated by two other
methods that
have been proposed to handle the scheme dependence.
The first method is the one used by the Particle Data Group.
The PDG uses $\overline{\textrm{MS}}$
and sets $\mu=\sqrt{s_{\textrm{\scriptsize CM}}}$. The second method that
we consider is the PMS-criterion. Information on this can be obtained
from~\cite{stevenson}. We also plot the result obtained by
our method of re-expanding in~$1/X_1$. We obtain figure~\ref{fig:hadrR}.
In this figure we did not yet use the symmetry of equations
(\ref{eq:Xsymm})~and~(\ref{eq:LXsymm}) to remove the logarithms.
If we use it, we get figure~\ref{fig:imphadrR}. The normalization that
we mentioned is done
by dividing by the one-loop result that can be found using the PDG-method.
This prevents the graphs from being very close to each others. We conclude
that our method gives results basically equal to the two other methods
provided that it is
improved by the use of the symmetry. ``Basically equal'' means that the
uncertainty that
results from ignoring the next order is much larger than the uncertainty
that comes from the renormalization scheme dependence.

\section{Comparison with Other Solutions}
Other solutions to the problem of renormalization scheme dependence have 
been proposed. The one that perhaps is most like ours, is by
C.J. Maxwell~\cite{maxwell}. In fact, in the case where~$\beta_1=0$ the
one-loop result of Maxwell is identical to ours. He sums some of the
higher order terms
along with the ones that come from orders where the full result is
known. If an $n$-loop
calculation has been performed, $X_n$ is known and all terms that
contain $X_i$ with $i\leq n$ are to be summed. This has the problem that
the definition of the $X_i$ depends on what parameters are chosen to 
parameterize the scheme. Maxwell notes this himself in~\cite{maxwell}.
As we have seen, the invariants are constants of integration, hence
they depend on the boundary chosen. Therefore it does not make much
sense to resum these, because it will inevitably lead to arbitrariness.
Our approach of expanding in the first invariant avoids this problem.

The PMS criterion, introduced in~\cite{stevenson} has the big advantage
that, apart from our own method, it is the only one that is really
completely scheme and convention independent. It has, however, the
disadvantage that it is difficult to apply. Here, the optimum value,
in some sense of optimal,
of scheme dependent parameters is determined. This involves
solving trancedental equations containing integrals, and hence can
generally only be done numerically. If one is interested in expressing
a physical quantity in, say, the numbers of flavours~$N_f$,
this can only be given as a set of equations. By contrast, we have a
series expansion in the parameter~$1/\tilde X_1$. Only one equation
needs to be solved to obtain $\tilde X_1$ from~$X_1$. Such a result can
easily be expressed in, for instance,~$N_f$.

The method of calculating hadronic~$R$ used by the Particle Data
Group consists of setting $\mu$ to a ``good''
value. This has the disadvantage
that one has to pick this ``good'' value alongside with a ``good''
renormalization scheme. For hadronic~$R$ the PDG makes $\mu$ equal
to the center of mass energy and uses $\overline{\textrm{MS}}$.

Another idea was put forward in~\cite{gupta}. Here we obtain a
differential equation for the physical quantity. We have
\beq\label{eq:gupta}
\frac{dR}{ds_{\textrm{\scriptsize CM}}}=f(R).
\eeq
As before, $s_{\textrm{\scriptsize CM}}$ is the energy squared in the
center of mass, not the scale. It turns out that the right-hand side
is scheme independent, which should not come as a surprise to the
reader of this paper. The method has some disadvantages compared to
ours. Firstly, this differential equation still needs to be solved.
This could be done by giving an initial value at some reference
energy, and then expanding
in~$\log(s_{\textrm{\scriptsize CM}}/s_{\textrm{\scriptsize ref}})$.
This, of course, goes wrong if the energy of an experiment
starts to differ significantly from the reference energy. Furthermore,
the mere mention of a ``reference energy'' indicates that
we are reintroducing arbitrariness. So, presumably, we are not to
solve this differential equation by expanding with respect to this
quantity. What are we to do then? Solving the
differential equation numerically, perhaps? The reader will not find
it difficult to think of disadvantages of this. Furthermore, we would
want to relate different physical quantities to each other. In this
method we would give a series expansion that
expresses one into the other. However, which of all possible physical
quantities is going to appear in a
listing of fundamental quantities? And at what energy is this
quantity going to be listed? This method has no preference for a
particular quantity or energy.

Yet another idea can be found in \cite{dhar}. Here an equation that
looks a lot like equation~(\ref{eq:running}) is given. In our notation
it would be given by
\beq\label{eq:effcharge}
X_1=-\frac{r_0^2}{R}
   +\frac{r_0\beta_1}{\beta_0}\log\frac{r_0\beta_0+\beta_1R}{\beta_1R}
   +\int_0^RdR'\,
		\left(\frac1{Y(R')}-\frac1{Y_0(R')^2+Y_1(R')^3}\right),
\eeq
where $Y(R)$ is a power series that starts with the term
$Y_0(R)^2$. The $Y_i$ are invariants related to the $X_i$ but their
definition is not entirely the same.
If we go to a scheme where all $r_i$ with $i>0$ are zero, the just given
equation and equation~(\ref{eq:running}) are
the same. Because only renormalization scheme invariants occur in
equation~\ref{eq:effcharge}, it must hold in any scheme. This could be an
answer to the question that bothered us in the previous paragraph,
because
this expression is a solution to the differential equation~(\ref{eq:gupta}).
The relation to our method is that we propose
to simplify this expression by turning it into an expansion.
Numerical accuracy is then achieved by resummation of logarithms.

The method proposed by \cite{grunberg} ultimately boils down to the same
equation~(\ref{eq:effcharge}). The philosophy is different though.
The idea is that the coupling constant should be interpreted as
an ``effective charge'' that no longer receives higher order corrections.

\section{Conclusions}
We can get rid of the unphysical dependence of physical quantities
on the renormalization scheme by expanding in~$1/X_1$ where $X_1$,
is the renormalization scheme invariant that occurs on the one loop
level. Numerical accuracy is achieved by resumming all logarithms
that contain~$X_1$ into the quantity $1/\tilde X_1$.

For future research it would be interesting to look at the possibility
of generalizing our approach to the case of a theory with masses and/or
multiple coupling constants. Another thing that could be done is to
try a similar approach to factorization scheme dependence as it arises
when one studies deep-inelastic scattering. In the context
of the method introduced in~\cite{maxwell}, both renormalzation and
factorization scale dependence are discussed in~\cite{fact}.


\begin{thebibliography}{9}
\bibitem{corr} K.G. Chetyrkin e.a., \textsl{QCD Corrections to the $e^+e^-$
Cross-section and the $Z$~Boson Decay Rate: Concepts and Results},
Phys.Rept.277:189-281, 1996.
\bibitem{bfun2} M. Czakon, \textsl{The four-loop QCD beta-function and
	anomalous dimensions}, hep-ph/0411261.
\bibitem{dhar} A. Dhar, \textsl{Renormalization Scheme-Invariant
	Perturbation Theory}, Phys. Lett. B128 (1983) 407.
\bibitem{grunberg} G. Grunberg, \textsl{Renormalization Group Improved
	Perturbative QCD},  Phys. Lett. B95 (1980) 70.
\bibitem{gupta} V. Gupta e.a., \textsl{New Perturbative Approach to
	General Renormalizable Quantum Field Theories},
	Int. J. Mod. Phys. A6: 3381-3398, 1991.
\bibitem{maxwell} C.J. Maxwell, \textsl{Complete Renormalization Group
	Im\-prove\-ment---\allowbreak Avoid\-ing Scale Dependence in QCD
	Predictions}, hep-ph/9908463, Nucl\@. Phys\@. Proc\@. Suppl\@.
	86:74-77,2000.
\bibitem{fact} C.J. Maxwell e.a., \textsl{Complete renormalization group
	improvement---avoiding factorization and renormalization scale
	dependence in QCD predictions}, hep-ph/0002204, Nucl. Phys. B577 (2000)
	209.
\bibitem{review} Particle Data Group, \textsl{Review of Particle Physics},
	Physics Letters B592 2004 1-1109, http://pdg.lbl.gov.
\bibitem{bfun} T. van Ritbergen e.a, \textsl{The four-loop
	$\beta$-function in Quantum Chromodynamics}, hep-ph/9701390,
	Phys.Lett. B400 (1997) 379-384.
\bibitem{stevenson} P.M. Stevenson, \textsl{Optimised Perturbation Theory},
	Phys. Rev. D23 (1981) 2916.
\end{thebibliography}
\end{document}